\documentclass[letters,fleqn,usenatbib,useAMS]{mnras}

\usepackage[T1]{fontenc}
\usepackage{ae,aecompl}

\usepackage{graphicx}	
\usepackage{amsmath}	
\usepackage{amssymb}	

\newcommand{\teff}{T_{\rm eff}}
\newcommand{\kms}{km\,s$^{-1}$}
\newcommand{\bz}{\langle B_z \rangle}
\newcommand{\nz}{\langle N_z \rangle}

\newcommand{\rsun}{$R_{\odot}$}

\title[The magnetic period of WR\,55]{The magnetic, spectroscopic, and photometric variability of the Wolf-Rayet star WR\,55}

\stepcounter{footnote}

\author[J\"arvinen et al.]{
S.~P.~J\"arvinen$^{1}$,
S.~Hubrig$^{1}$,
R.~Jayaraman$^{2}$,
A.~Cikota$^{3}$,
M.~Sch\"oller$^{4}$
\\
$^{1}${Leibniz-Institut f\"ur Astrophysik Potsdam (AIP), An der Sternwarte~16, 14482~Potsdam, Germany} \\
$^{2}${MIT Kavli Institute and Department of Physics, 77 Massachusetts Avenue, Cambridge, MA 02139, USA}\\
$^{3}${Gemini Observatory / NSF’s NOIRLab, Casilla 603, La Serena, Chile} \\
$^{4}${European Southern Observatory, Karl-Schwarzschild-Str.~2, 85748 Garching, Germany}
}

\date{Accepted 2023 June 02. Received 2023 May 22; in original form 2023 April 06}

\pubyear{2023}

\begin{document}
\label{firstpage}
\pagerange{\pageref{firstpage}--\pageref{lastpage}}
\maketitle

\begin{abstract}
Studies of magnetic fields in the most evolved massive stars, the Wolf-Rayet 
stars, are of special importance because they are progenitors of certain 
types of supernovae. The first detection of a magnetic field of the order of 
a few hundred Gauss in the WN7 star WR\,55, based on a few FORS\,2 
low-resolution spectropolarimetric observations, was reported in 2020. 
In this work we present new FORS\,2 observations allowing us to 
detect magnetic and spectroscopic variability with a period of 11.90\,h. 
No significant frequencies were detected in TESS and ASAS-SN photometric 
observations. Importantly, magnetic field detections are achieved currently 
only in two Wolf-Rayet stars, WR\,6 and WR\,55, both showing the presence of
corotating interacting regions.
\end{abstract}




\section{Introduction}

The role of magnetic fields in the evolution of massive stars remains poorly 
understood. Massive O stars have been found to have large-scale organized, 
predominantly dipolar magnetic fields
\citep[e.g.,][]{Hubrig2013, Grunhut, BOB, Hubrig2023}, 
so that we expect to detect such fields in their descendants, the Wolf-Rayet 
(WR) stars.
\citet{Aguilera} 
calculated a sequence of rotating, magnetized pre-collapse models for WR stars 
with sub-solar metallicity and suggested that the presence of magnetic fields 
could explain extreme events such as superluminous supernovae and gamma-ray 
bursts powered by proto-magnetars or collapsars. They concluded that further
modelling of WR stars with different metallicities and rotational velocities 
will allow for comparisons with observed rates of superluminous supernovae 
and gamma-ray bursts.

\citet{GayleyIgnace} 
predicted a fractional circular polarization of a few times $10^{-4}$ for 
WR magnetic fields of about 100\,G. Verifying this prediction 
observationally, however, is difficult due to the WR stars' significantly 
broadened emission-line spectra arising from a strong stellar wind. Recent 
work has found evidence for WR magnetic fields: marginal detections in three 
stars by
\citet{delacherv}, 
who used high-resolution spectropolarimetry, and a 3.3$\sigma$  detection of 
a mean longitudinal magnetic field $\bz$ in the cyclically variable, X-ray 
emitting WN5-type star WR\,6 by 
\citet{Hubrig2016}, 
who used low-resolution FORS\,2 spectropolarimetry 
\citep{Appenzeller1998}. 
Cyclical variability can arise from a corotating interacting region (CIR), 
whose signatures include different photospheric absorption features 
simultaneously propagating with different accelerations. CIRs in massive 
stars have been found using spectroscopic time series
\citep[e.g.,][]{Mullan}. 
These may relate to magnetic bright spots, which often signal the presence 
of a global magnetic field 
\citep[e.g.,][]{Ramiaramanantsoa}.

\citet{Hubrig2020} 
reported a definite detection in another WR star with similar cyclical 
variability
\citep{Chene-St},
the hydrogen-deficient WN7-type star WR 55 
\citep[$\teff$ = 56.2\,kK;][]{Hamann2006}. 
The measured $\bz$ showed a change in polarity, with the highest field 
values $\bz=-378\pm85$\,G (4.4$\sigma$) and $\bz=205\pm58$\,G (3.5$\sigma$).
This work presents additional FORS\,2 spectropolarimetric observations of 
WR\,55, alongside photometry, and aims to constrain the magnetic field 
strength and identify any periodicities.

\section{Magnetic and spectral variability}
\label{sec:specpol}

Twelve new FORS\,2 spectropolarimetric observations, in addition to the four 
observations presented in 
\citet{Hubrig2020}, 
are listed in Table~\ref{tab:obs}. The new spectra were acquired between 2022 
February 16 and April 11, with the same instrument setup as in 
\citet{Hubrig2020}.
The wavelength was calibrated with a He-Ne-Ar arc lamp, and the extraction 
of the ordinary and extraordinary beams was done using standard {\sc iraf} 
procedures 
\citep{Cikota}. 
The measurements of the mean longitudinal magnetic field were carried out 
using procedures presented in prior work
\citep[e.g.,][and references therein]{Hubrig2004a, Hubrig2004b, BOB}.

\begin{table}
\caption{
Longitudinal magnetic field values $\bz$ of WR\,55 from FORS\,2 
spectropolarimetric observations. The first column gives the modified Julian 
date (MJD) of mid-exposure, followed by the corresponding rotational phase 
for a period of 11.90\,h and $T_0=59664.226$, and then the values for the 
signal-to-noise ratio (S/N) of the Stokes~$I$ spectra measured near 
4686\,\AA{}. The last two columns show the $\bz$ measurements from the Monte 
Carlo bootstrapping test, and the corresponding measurements $\nz$ from the 
null spectra. Errors represent 1$\sigma$ uncertainties. Observations with * 
are from 
\citet{Hubrig2020}.
}
\label{tab:obs}
\centering
\begin{tabular}{lcr r@{$\pm$}l r@{$\pm$}l}
\hline
\multicolumn{1}{l}{MJD} &
\multicolumn{1}{c}{$\varphi$} &
\multicolumn{1}{c}{S/N} &
\multicolumn{2}{c}{$\bz$} &
\multicolumn{2}{c}{$\nz$} \\
\multicolumn{1}{l}{} &
\multicolumn{1}{r}{} &
\multicolumn{1}{r}{} &
\multicolumn{2}{c}{G} &
\multicolumn{2}{c}{(G)} \\
\hline
58892.2058* & 0.067 & 2388 & 205    & 58 & $-$16 & 55 \\
58898.3492* & 0.457 & 2579 & $-$378 & 85 & $-$6  & 78 \\
58900.2075* & 0.204 & 2168 & 56     & 80 & 63    & 82 \\
58916.3059* & 0.670 & 2386 & 4      & 66 & $-$16 & 57 \\
59627.2359  & 0.401 & 3402 & $-$172 & 47 & $-$26 & 45 \\
59652.2534  & 0.854 & 2618 & 143    & 81 & 29    & 79 \\
59660.0977  & 0.674 & 2726 & $-$47  & 49 & $-$37 & 48 \\
59662.3518  & 0.219 & 3057 & $-$73  & 38 & $-$16 & 41 \\
59664.3504  & 0.250 & 3073 & $-$33  & 34 & 18    & 31 \\
59666.3152  & 0.212 & 2822 & $-$35  & 55 & 20    & 46 \\
59667.2918  & 0.182 & 3082 & $-$129 & 42 & $-$13 & 46 \\
59668.3296  & 0.275 & 2599 & $-$44  & 52 & 38    & 46 \\
59670.0844  & 0.814 & 2710 & 114    & 48 & $-$19 & 49 \\
59671.2429  & 0.152 & 2755 & 81     & 47 & $-$25 & 48 \\
59679.2081  & 0.213 & 2818 & $-$26  & 43 & $-$24 & 42 \\
59681.0112  & 0.850 & 2455 & 156    & 53 & 54    & 55 \\
\hline
\end{tabular}
\end{table}

\begin{figure}
\centering
\includegraphics[width=0.44\textwidth]{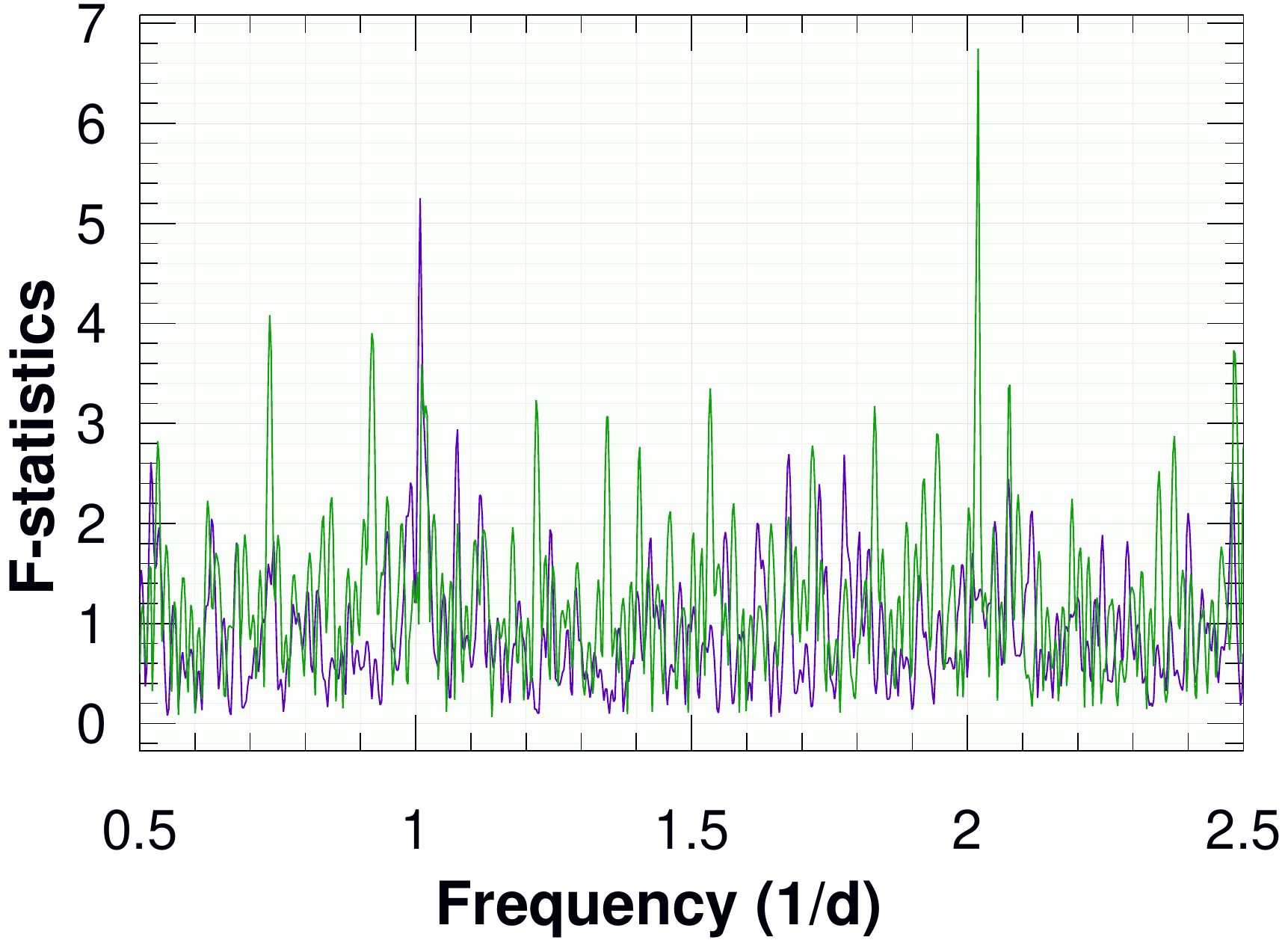}
\includegraphics[width=0.4\textwidth]{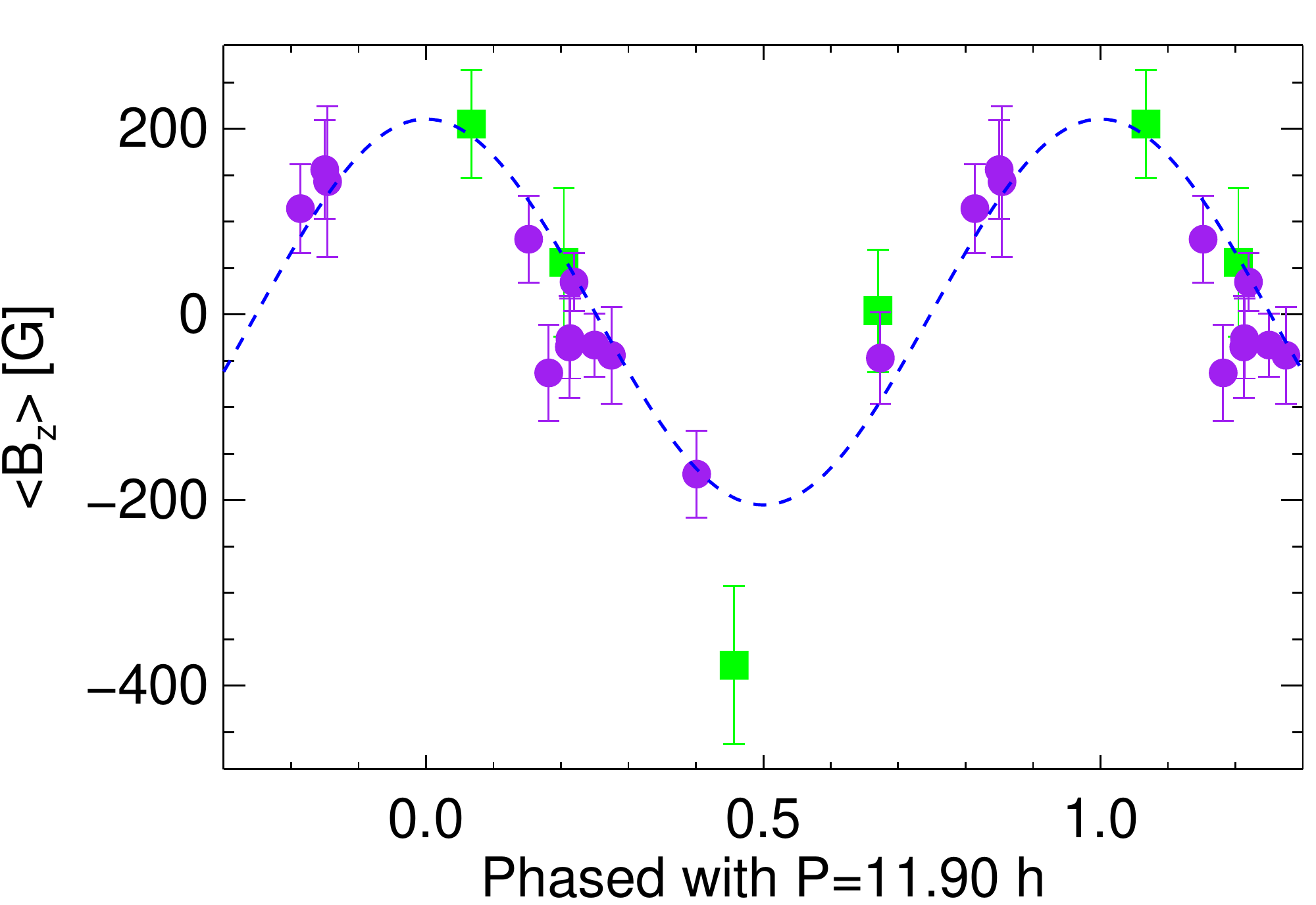}
\caption{
\emph{Top:}
F-statistics periodogram (green) for the $\bz$ measurements of WR\,55, with 
the window function (purple).
\emph{Bottom:}
$\bz$ measurements of WR\,55 from low-resolution FORS\,2 spectropolarimetric 
observations, phased with the period of 11.90\,h. The dashed blue curve 
corresponds to the sinusoidal fit. Green squares are measurements from 
\citet{Hubrig2020}, 
purple dots are from 2022. Vertical bars indicate the measurement accuracy.
}
\label{fig:Bz}
\end{figure}

Our frequency analysis based on the $\bz$ measurements in 
Table~\ref{tab:obs} was performed using a Levenberg-Marquadt non-linear 
least-squares fit
\citep{press}.
To detect the most probable period, we calculated the frequency spectrum, 
and for each trial frequency, we performed a statistical F-test of the null 
hypothesis -- the absence of periodicity
\citep{seber}. 
The resulting F-statistic can be thought of as the total sum, including 
covariances, of the ratio of harmonic amplitudes to their standard 
deviations, i.e., a S/N. As shown in the top panel of Fig.~\ref{fig:Bz}, 
the highest peak in the frequency spectrum corresponds to a period 
$P=11.9006\pm0.0001$\,h. The distribution of the measured $\bz$ values over 
this period assuming the ephemeris $T_0=59664.2260$ is presented in the 
bottom panel of Fig.~\ref{fig:Bz}. Assuming a dipolar field structure, this 
ephemeris corresponds to the maximum positive field extremum in the 
corresponding sinusoidal field phase curve.

\begin{figure*}
\centering
\includegraphics[width=0.27\textwidth]{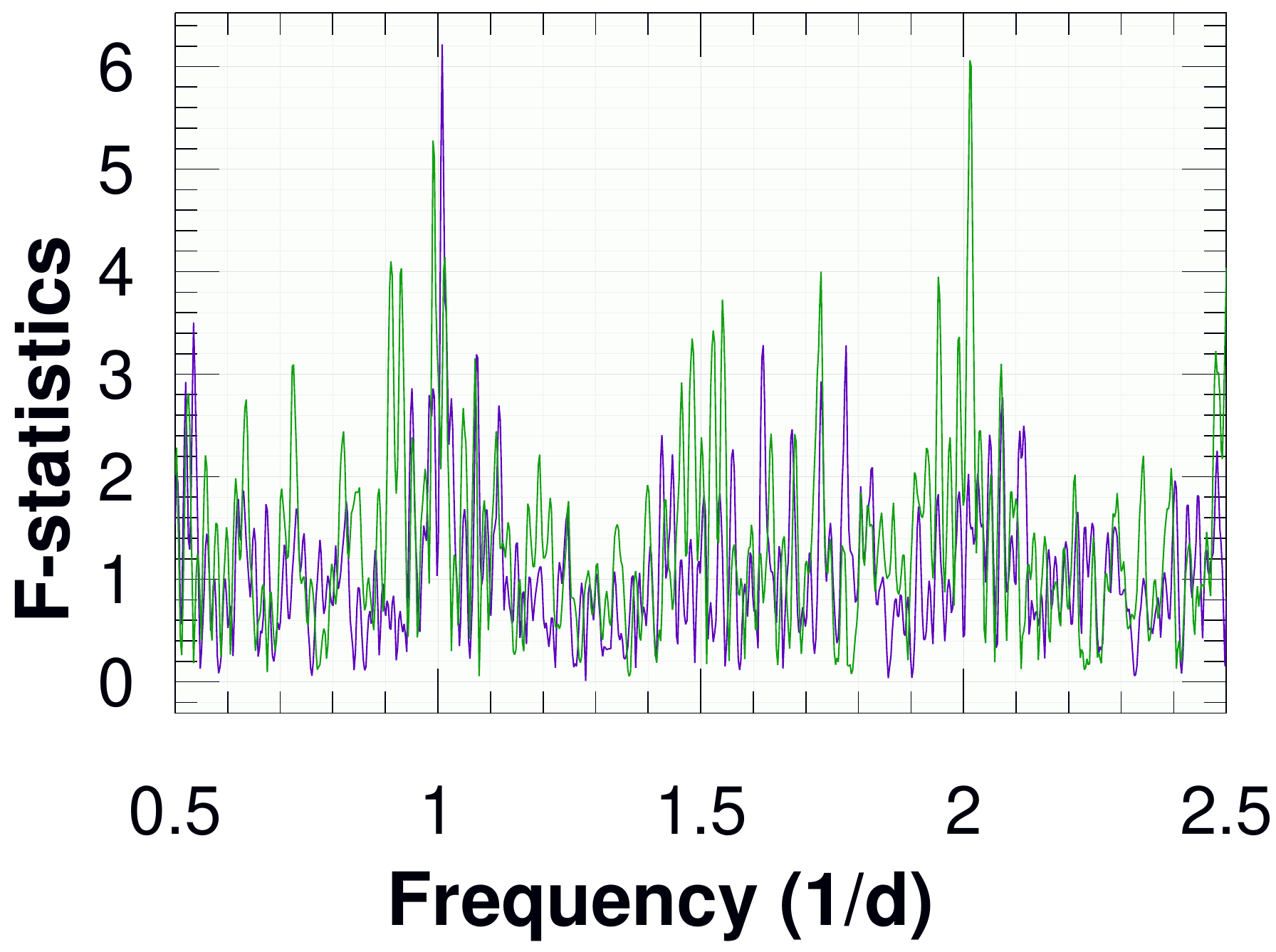}
\includegraphics[width=0.27\textwidth]{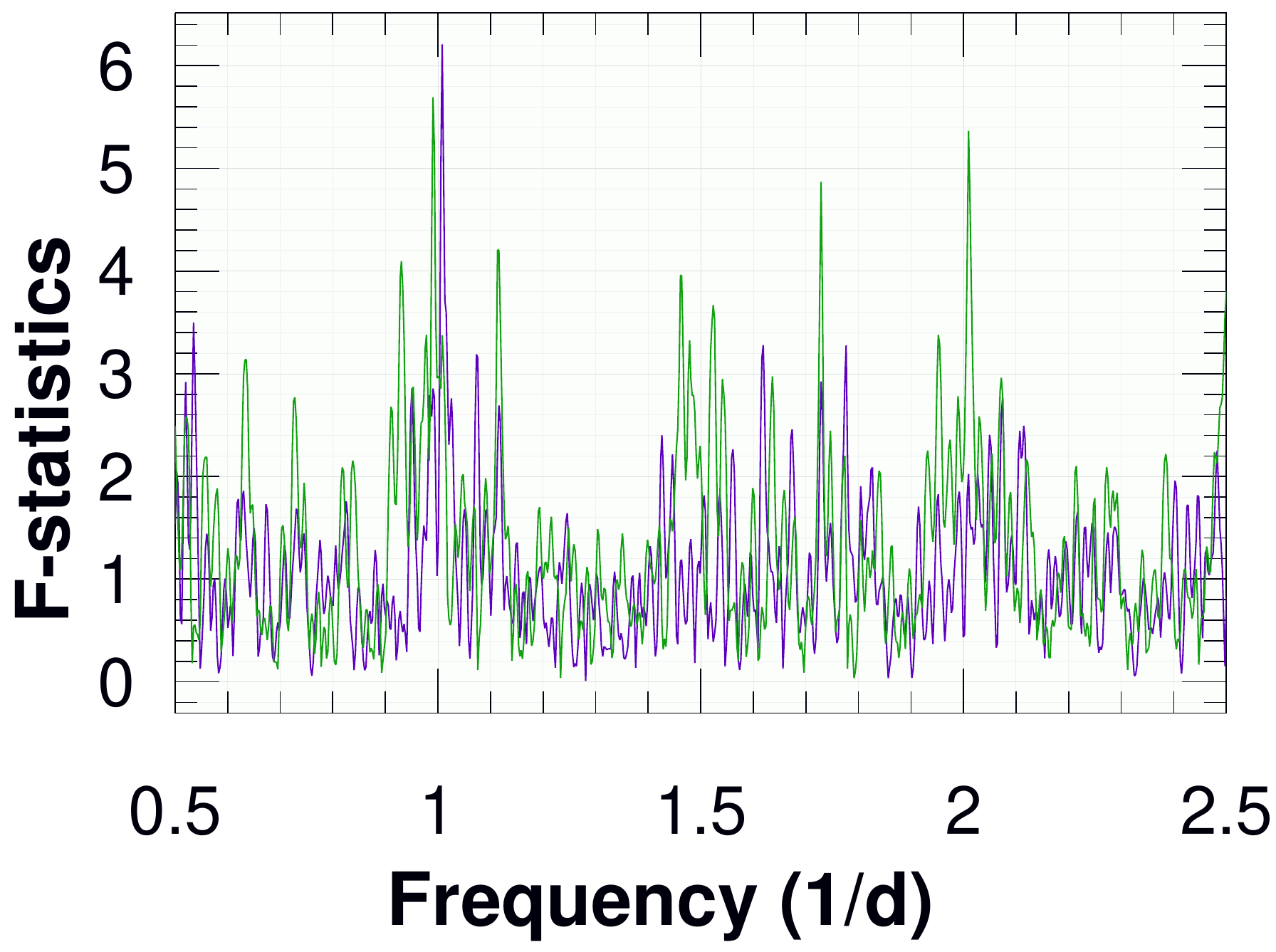}
\includegraphics[width=0.27\textwidth]{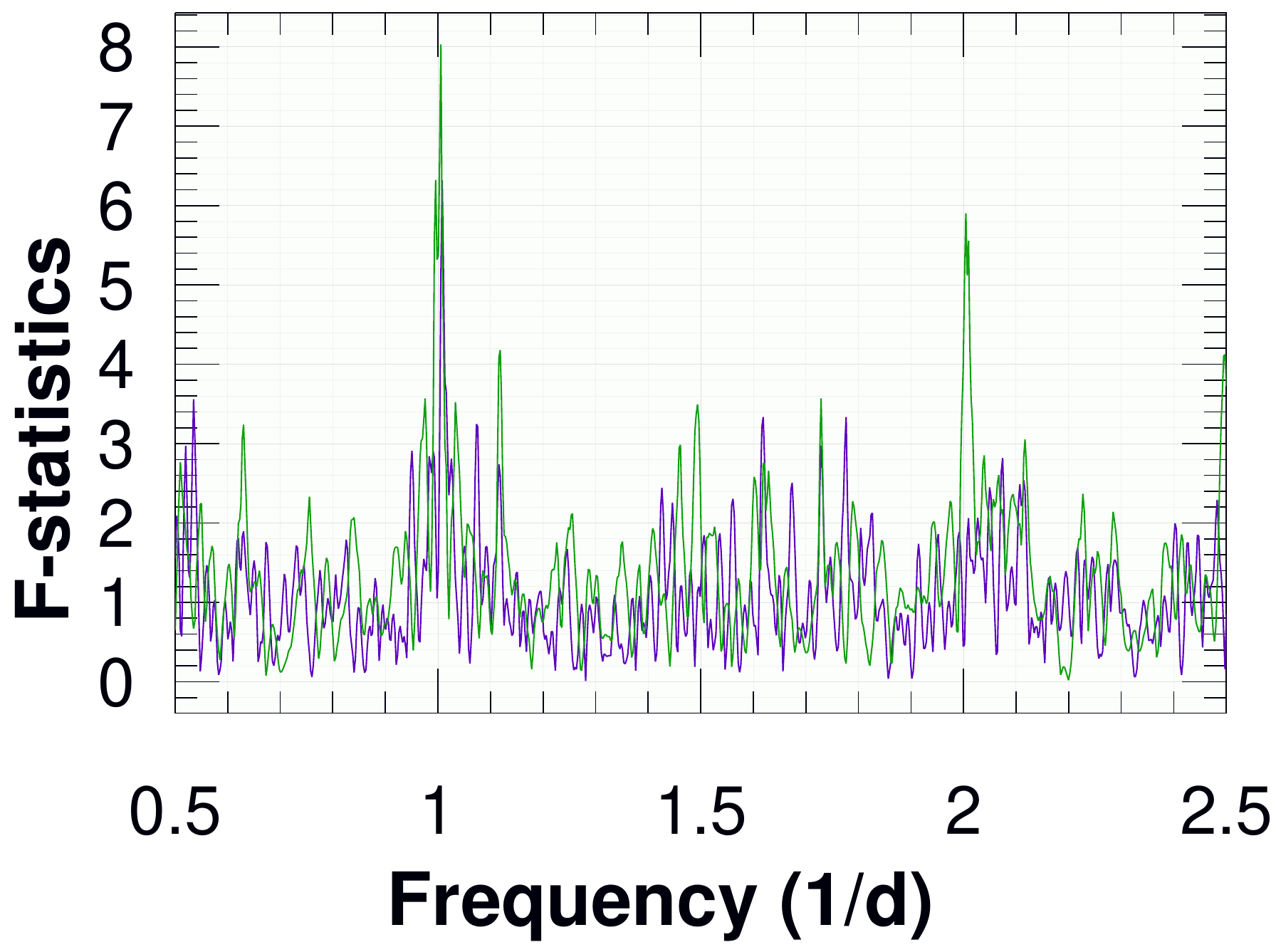}
\includegraphics[width=0.27\textwidth]{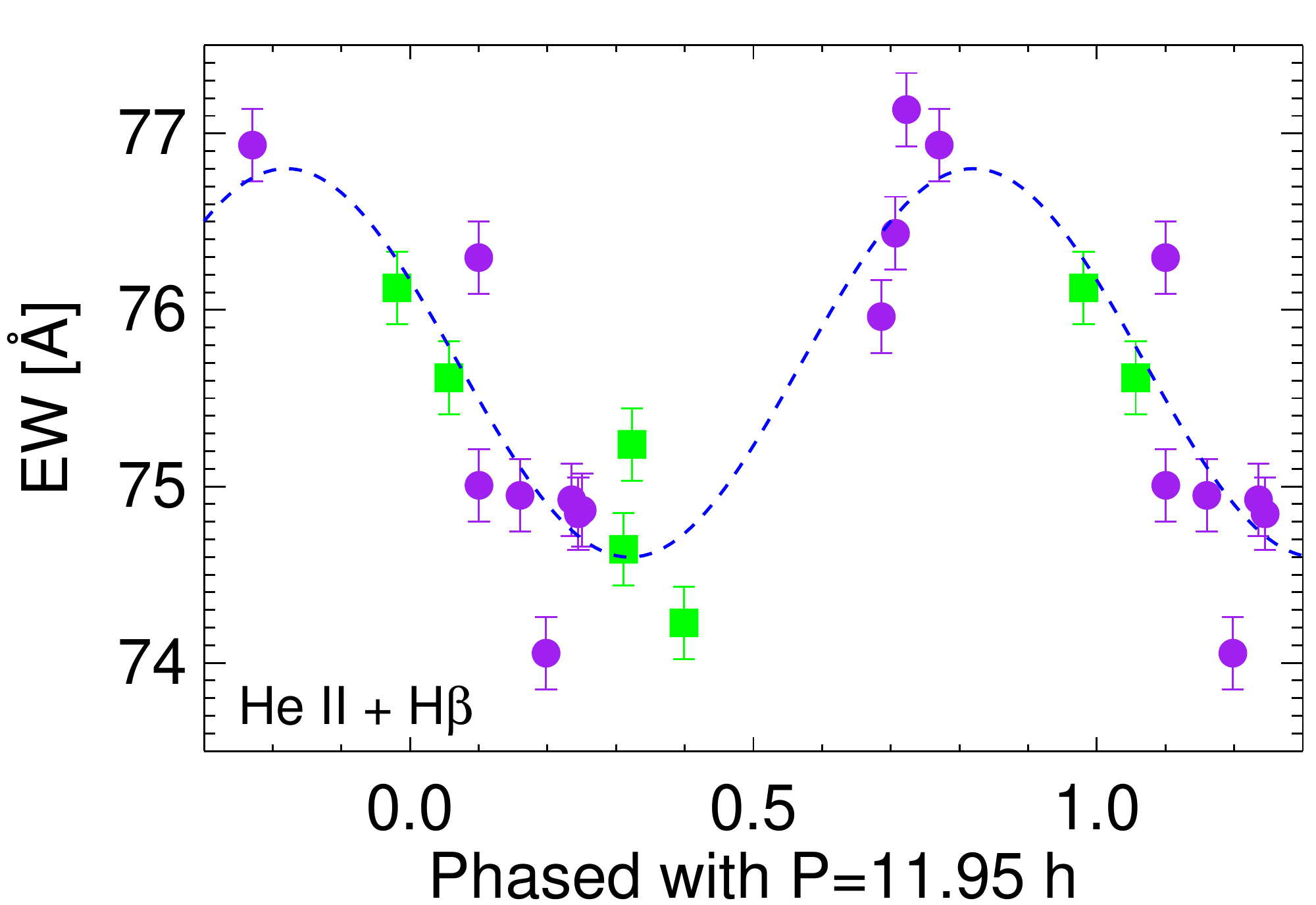}
\includegraphics[width=0.27\textwidth]{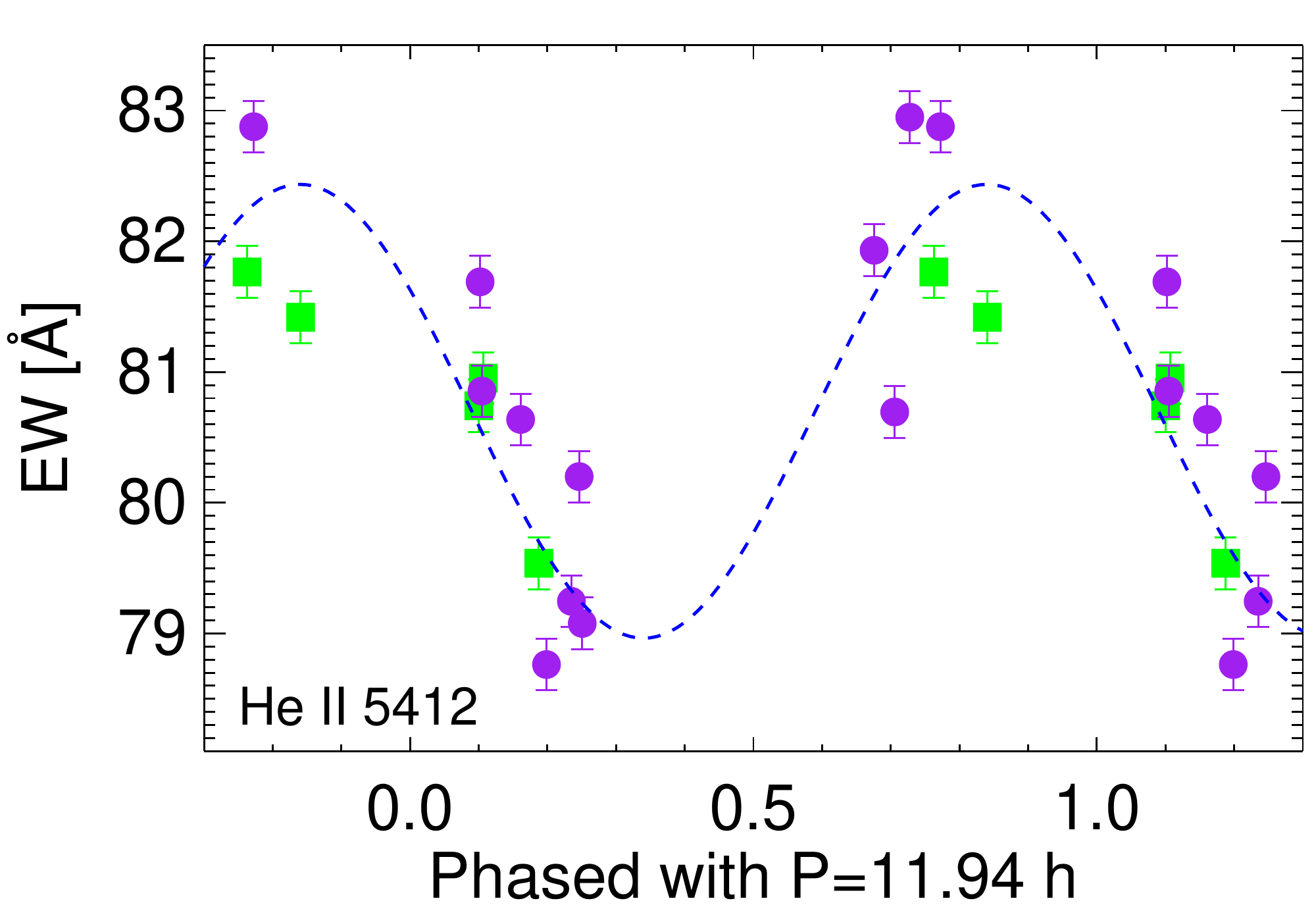}
\includegraphics[width=0.27\textwidth]{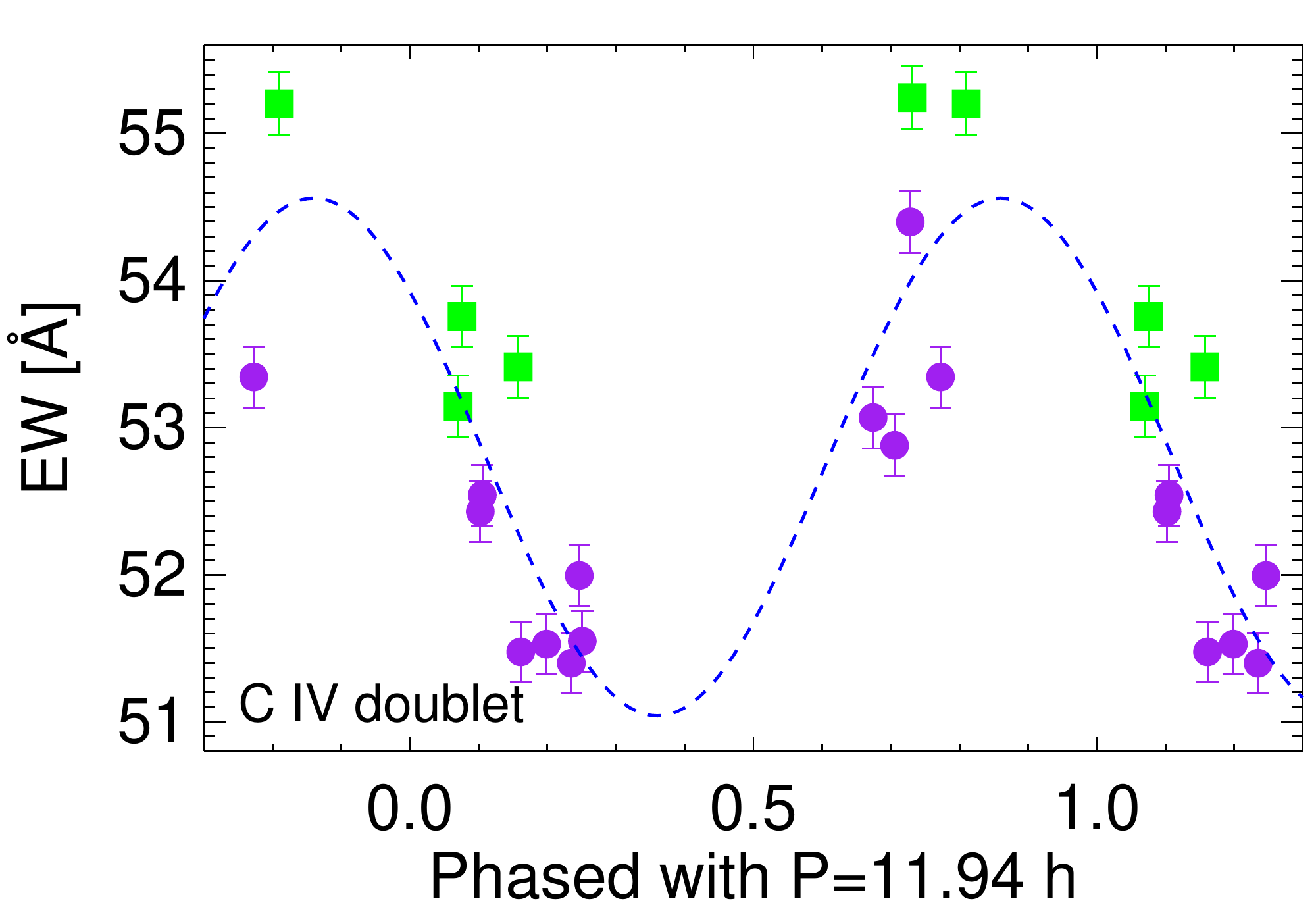}
\caption{
\emph{Top:}
F-statistics periodogram (green) for equivalent widths of the \ion{He}{ii}
and H$\beta$ blend (left), the \ion{He}{ii}~5412\,\AA{} line (middle), and
the \ion{C}{iv} doublet (right).
\emph{Bottom:}
The corresponding phase curves, with symbols as in Fig.~\ref{fig:Bz}.
Vertical bars show measurement accuracy.
}
\label{fig:EWs}
\end{figure*}

Additionally, three spectral lines, the \ion{He}{ii} and H$\beta$ blend, the 
\ion{He}{ii}~5412\,\AA{} line, and the \ion{C}{iv} doublet were investigated 
for the presence of rotationally-modulated variability of line intensities 
and radial velocities. The results of the frequency analysis of equivalent 
widths (EWs) for all three lines are presented in the top panel of 
Fig.~\ref{fig:EWs}. The highest peaks in the frequency spectra not 
coinciding with the window function correspond to slightly longer periods:
$P=11.95\pm0.0020$\,h for the \ion{He}{ii}/H$\beta$ blend, and 
$P=11.94\pm0.0002$\,h for the \ion{He}{ii} and \ion{C}{iv} lines. Notably,
our frequency analysis indicates a lower significance for the periods 
obtained using EWs than for the period using the magnetic field measurements 
directly: the reduced $\chi^2$-values for the fits are 1.10 for the $\bz$ 
measurements, 5.83 for the \ion{He}{ii}/H$\beta$ blend, 15.00 for the 
\ion{He}{ii}~5412\,\AA{} line, and 13.30 for the \ion{C}{iv} doublet. The 
distributions of the EW values over the detected periods, assuming the same 
ephemeris as for the magnetic field measurements, are presented in the 
bottom panel of Fig.~\ref{fig:EWs}. We observe a clear phase shift of 0.22 
between the maxima of the measured EW and the magnetic maximum.
As WR\,55 is a CIR-type variable target, a curvature due to a spiral CIR 
can result in the development of a potentially exploitable phase lag
\citep{Ignace}.

Variability of EWs can be caused either by an inhomogeneous distribution of 
temperature or chemical elements on the surface of a rotating star, or a 
companion. Concerning the last possibility, we find no information in the 
literature on the binarity of WR\,55. We do not detect any significant 
frequency peaks in our measurements of the radial velocities. This may be 
because the emission-line spectra of WR stars are mainly formed in expanding 
atmospheric layers, where the wavelength and shapes of emission lines have 
multiple contributions: from wind velocity, stellar rotation, and 
(potential) orbital motion. Because CIRs may be explained via a paradigm 
involving bright spots driving CIR structures, we assume that the observed 
EW variability can be explained by temperature spots. The individual lines 
used in the equivalent widths measurements are most likely formed slightly 
higher up in the expanding atmosphere (we estimate by 0.3\% of the stellar 
radius), so that the rotation period becomes longer due to the conservation 
of angular momentum.

\section{Photometric variability}

\subsection{TESS photometry and contamination}

WR\,55 was observed by TESS in Sectors 11 and 38, from 2019 April 22 to
2019 May 21, and from 2021 April 28 to 2021 May 26. We constructed custom light
curves for Sector 11 with {\tt TESSCut}
\citep{2019ascl.soft05007B},
and for Sector 38 from the Target Pixel File (TPF) output by the Science
Processing Operations Center (SPOC) pipeline
\citep{jenkins-spoc},
using {\tt lightkurve}
\citep{2018ascl.soft12013L}.

TESS light curves often suffer significant contamination from nearby stars
due to the large plate scale (21''/pixel). For WR\,55, there is a nearby
bright ($G_{\rm RP} = 8.63$) $\beta$\,Cephei variable star (HD\,117704)
whose flux could fall into the chosen aperture.
Figure~\ref{fig:tess_lc_red_noise} shows the magnitudes (in G$_{\rm RP}$
band, which spans a similar range as the TESS passband) and locations of
bright (G$_{\rm RP} < 13$) sources near WR\,55, from Gaia Data Release 3
\citep{gaia-mission, gaia-dr3}.
To mitigate possible contamination, we used a custom one-pixel aperture
including only WR55 (light green shading in the left panel of
Fig.~\ref{fig:tess_lc_red_noise}; purple shading corresponds to the SPOC
aperture). The 30-minute cadence light curve from Sector~11 was extracted
via {\tt TESSCut} using a similar one-pixel aperture; backgrounds were
estimated as in section~2 of
\citet{toala-wr7}.
These light curves did not need detrending. The light curves and periodograms
of Sectors~11 and 38 are similar. 

\subsection{The Low-Frequency Excess}
\label{sec:red_noise_fitting}

We used the Discrete Fourier Transform 
\citep{deeming-dft} 
to calculate the periodograms of the light curves.
\citet{bowman-blue-sg,bowman-2019}, 
\citet{naze-2021}, 
and 
\citet{lenoir-craig-2022} 
assign a physical interpretation to the low-frequency excess (red noise) in 
the frequency spectra of massive stars. For WR stars in particular, three 
explanations have been posited: (a) interactions between clumped stellar 
wind and pulsations 
\citep{naze-2021}, 
(b) internal gravity waves 
\citep{bowman-2019}, 
and (c) a sub-surface convection zone caused by an iron-peak opacity bump 
\citep[``FeCZ''; see, e.g.,][]{canetiello-fecz}, 
which is most pronounced in WR stars. 
\citet{lenoir-craig-2022}
also suggest a line-deshadowing instability (LDI) to explain the variability.
Finally, CIRs can also induce flux variability, but this was studied only in 
the radio 
\citep{Ignace}.

\begin{figure*}
\centering
\includegraphics[width=.9\textwidth]{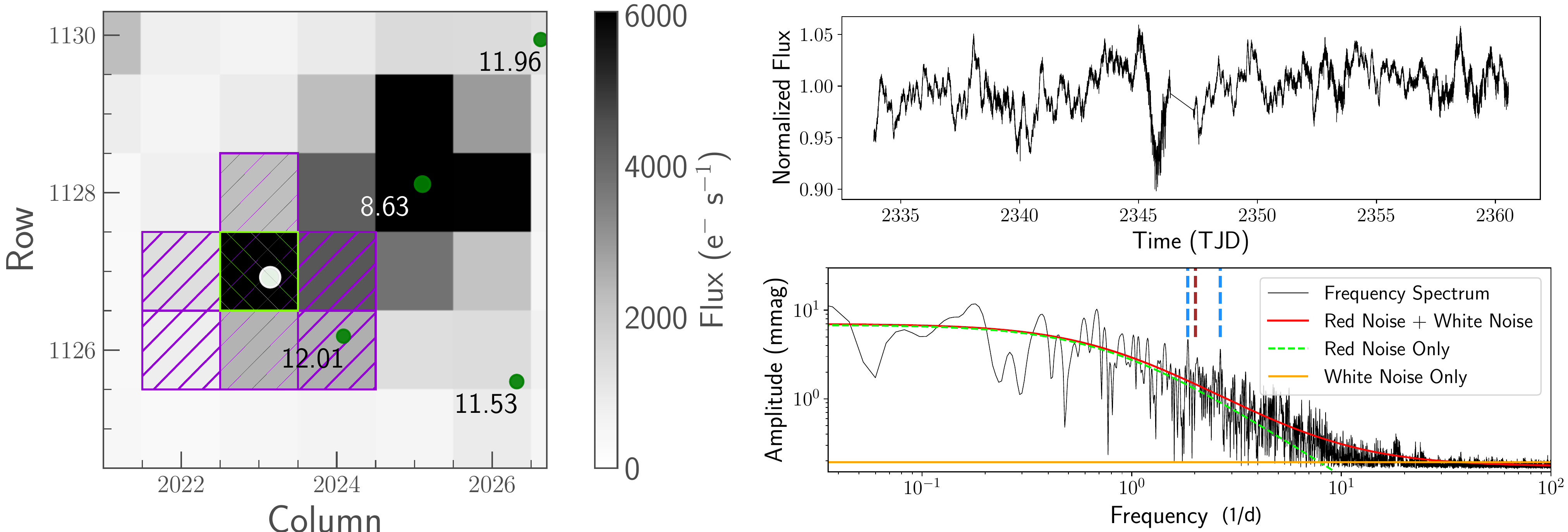}
\caption{
\emph{Left}:
The Sector~38 target pixel file of WR\,55, with nearby Gaia sources 
(G$_{\rm RP} < 13$) and magnitudes. WR\,55 is in white; other sources are in 
dark green. The $\beta$\,Cephei star HD\,117704, with $G_{\rm RP}$ = 8.63 
source, is labeled in white. Purple shading represents the photometric 
aperture chosen by the SPOC pipeline
\citep{jenkins-spoc};
light green represents our chosen aperture for the custom light curve
in the top right panel.
\emph{Right}:
The bottom panel shows the best-fit semi-Lorentzians for the power spectrum 
for the light curve in the top panel, based on the parameters in 
Table~\ref{tab:fit_params}. The black line is the power spectrum, while the 
red line represents the best-fit red + white noise model. These two are also 
plotted separately (light green and orange, respectively). Marginal 
frequencies are denoted by blue dotted lines (see further discussion in the 
text), while purple corresponds to the magnetic period obtained in 
Section~\ref{sec:specpol}.}
\label{fig:tess_lc_red_noise}
\end{figure*}

\begin{figure}
\centering
\includegraphics[width=.8\linewidth]{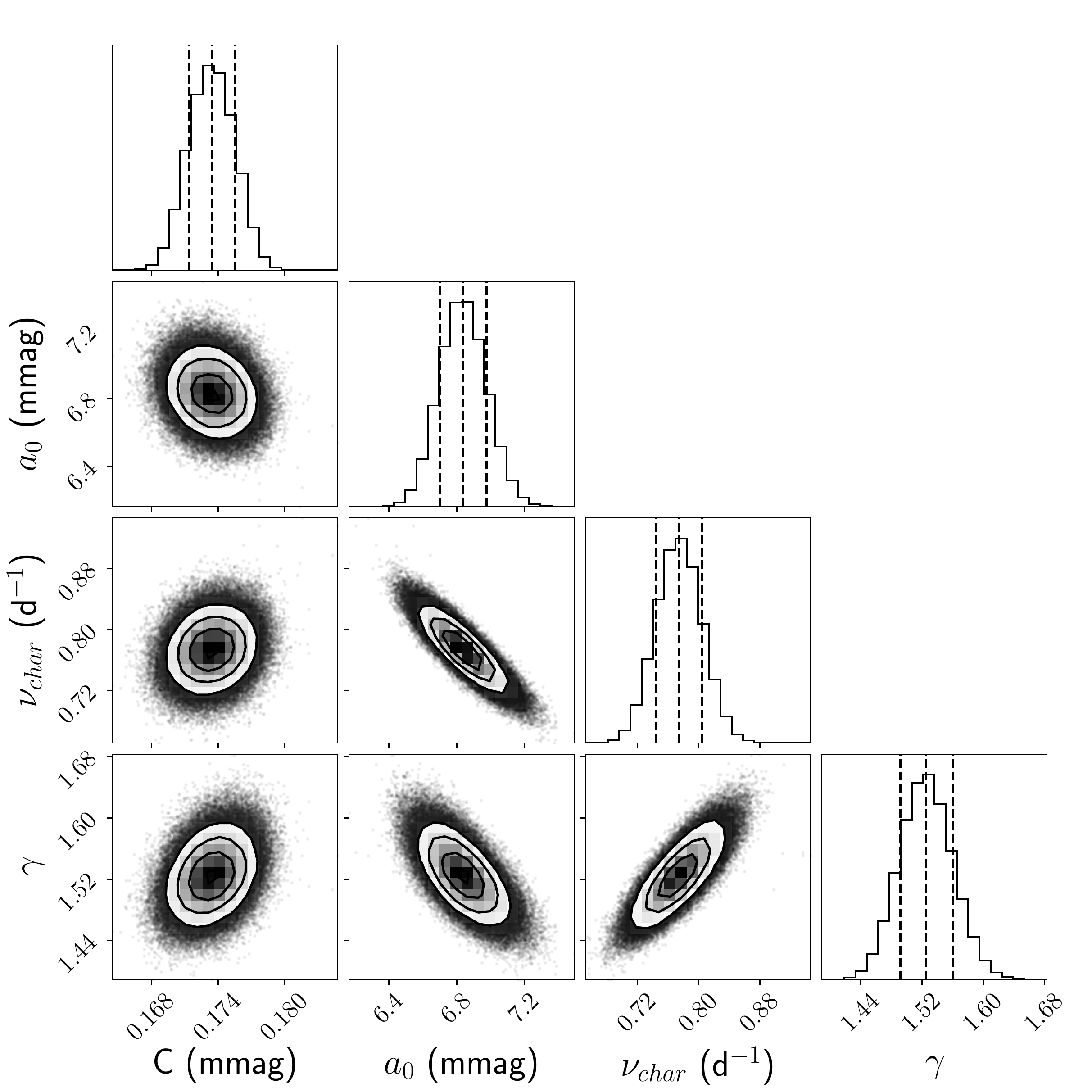}
\caption{
Posterior distributions for Sector 38 red noise parameters of WR\,55. Dashed 
vertical lines represent the 16th, 50th, and 84th percentiles (left to 
right). Strong negative correlations exist for ($a_0$, $\nu_{\rm char}$), 
($a_0$, $\gamma$), and a strong positive one for ($\nu_{\rm char}$, $\gamma$).
}
\label{fig:posterior_plot}
\end{figure}

As in prior literature, we fit the red noise with a semi-Lorentzian function 
added to a white noise term:
\begin{equation}
A(\nu) = \frac{A_0}{1 + \left( \frac{\nu}{\nu_{\rm char}}\right)^{\gamma}} + C_W,
\end{equation}
where $A_0$ represents the amplitude scale of the low-frequency variability,
$\nu_{\rm char}$ is its characteristic frequency, $\gamma$ is an exponent, 
and $C_W$ is the white noise component, which also captures instrumental 
variability. For our fit, we compare a custom implementation of the Markov 
Chain Monte Carlo (MCMC) method and the Levenberg-Marquadt non-linear 
least-squares method
\citep[with {\tt lmfit};][]{newville-2014-lmfit}.

To calculate formal uncertainties in power spectra amplitudes, we first
assumed a Poisson uncertainty ($\sqrt{N}$;  N is the count of 
e$^{-}$/second) for each point in both light curves and then bootstrapped 
1000 realizations of these points, with Gaussian noise. We then calculated 
the Lomb-Scargle periodogram using {\tt lightkurve}
\citep{1976Ap&SS..39..447L, 1982ApJ...263..835S, 2018ascl.soft12013L} 
and then used the standard deviation of the amplitudes in each frequency 
bin (across all 1000 power spectra) as the formal per-point uncertainty to be 
used when fitting, in order to derive parameter uncertainties. We contrast 
this with
\citet{naze-2021}, 
wherein parameter uncertainties are the square root of the diagonal of the 
covariance matrix, multiplied by $\sqrt{\chi^2_{\rm best\ fit}}$, divided 
by $0.5 \times N_{\rm data} - 4$ (degrees of freedom).

The MCMC ran for $5\times 10^6$ steps; priors (see 
Table~\ref{tab:fit_params}) were derived iteratively, with initial 
estimates based on the ranges of values in
\citet{naze-2021} 
and 
\citet{lenoir-craig-2022}. 
Best-fit parameter values from both techniques are in 
Table~\ref{tab:fit_params}; they agree to within their (similar) mutual 
uncertainties. Posterior distributions for Sector 38 parameters are in 
Fig.~\ref{fig:posterior_plot}. This is the first application and comparison 
of both techniques for the same star, showing that they are equally valid;
\citet{lenoir-craig-2022} 
solely used MCMC, while 
\citet{naze-2021} 
used only Levenberg-Marquadt. The best-fit semi-Lorentzians are shown in 
Fig.~\ref{fig:tess_lc_red_noise}; an F-test shows that a red noise model is 
strongly favored over a white-noise-only model ($p < 10^{-16}$).

\begin{table*}
\caption{
Best-fit parameters and uncertainties from both fitting methods for all
data for WR 55, along with MCMC priors. 
}
\label{tab:fit_params}
\centering
\begin{tabular}{ccccccc}
\hline
\multicolumn{1}{c}{} &
\multicolumn{4}{c}{MCMC} &
\multicolumn{2}{c}{Least-Squares} \\
\hline
\multicolumn{1}{c}{Parameter} &
\multicolumn{1}{c}{Sec. 11} &
\multicolumn{1}{c}{Sec. 11 Prior} &
\multicolumn{1}{c}{Sec. 38} &
\multicolumn{1}{c}{Sec. 38 Prior} &
\multicolumn{1}{c}{Sec. 11} &
\multicolumn{1}{c}{Sec. 38}  \\
\hline
$C_W$ (mmag) & 0.697\,$\pm$\,0.012 & 0.6\,$\leq\,C_W\,\leq$\,0.8 & 0.173\,$\pm$\,0.002 & 0.15\,$\leq\,C_W\,\leq$\,0.2 & 0.697\,$\pm$\,0.014 & 0.173\,$\pm$\,0.001 \\
$A_0$ (mmag) & 10.9\,$\pm$\,0.2 & 10\,$\leq\,A_0\,\leq$\,13 & 6.84\,$\pm$\,0.14 & 5\,$\leq\,A_0\,\leq$\,8 & 11.0\,$\pm$\,0.3 & 6.84\,$\pm$\,0.04\\
$\nu_{\rm char}$ (d$^{-1}$) & 0.300\,$\pm$\,0.012 & 0.25\,$\leq\,\nu_{\rm char}\,\leq\,$0.35 & 0.774\,$\pm$\,0.030 & 0.6\,$\leq\,\nu_{\rm char}\,\leq$\,1.0 & 0.300\,$\pm$\,0.014 & 0.772\,$\pm$\,0.007 \\
$\gamma$ & 1.39\,$\pm$0.03 & 1.25\,$\leq\,\gamma\,\leq\,$1.5 & 1.53$^{+0.04}_{-0.03}$ & 1.25\,$\leq\,\gamma\,\leq\,$1.75 & 1.38\,$\pm$\,0.04 & 1.52\,$\pm$\,0.01\\
\hline
\end{tabular}
\end{table*} 

\subsection{Photometric Peaks}

No significant (S/N $> 5$) or marginal (3 $<$ S/N $<$ 5) peaks were found in 
Sector\,11. In Sector\,38, we find two marginal peaks at 
1.8544\,$\pm$\,0.0003\,d$^{-1}$ and 2.6485\,$\pm$\,0.0003\,d$^{-1}$ (blue 
dashed lines in Fig.~\ref{fig:tess_lc_red_noise}). To search for 
longer-period variability, we used the light curve from the All-Sky 
Automated Survey for Supernovae Sky Patrol web server
\citep[][with a Nyquist limit of $\sim$0.5\,d]{asassn-lc}. 
Data were separated by camera, and the periodograms were calculated with 
{\tt astropy}
\citep{2013A&A...558A..33A}. 
We found marginal peaks at $\sim$6.4\,d (in camera {\tt be}) and 
$\sim$20--25 d$^{-1}$ (in both cameras); we do not ascribe physical meaning 
to these.

None of these periods matches $P = 11.9$\,h from Sect.~\ref{sec:specpol}
(the brown dashed line in the bottom right panel of
Fig.~\ref{fig:tess_lc_red_noise}), or the observed pulsation frequencies
of the nearby $\beta$\,Cep star. Marginal peaks may correspond to pulsation
modes, which shift, appear, and disappear over many months, and may underlie
the discrepancy in red noise parameters between sectors. Recovery of
WR\,55's photometric rotation period may also be hindered by circumstellar
material.

\section{Discussion and Conclusions}
\label{sec:disc}

\citet{Hubrig2020} 
found a magnetic field in WR\,55, and our additional observations confirm 
this and constrain the $\bz$ amplitude to $\sim$ 200\,G. The sinusoidal 
field phase curve indicates a likely dipolar field structure. After the 
detection of a longitudinal magnetic field in WR\,6, WR\,55 is now the 
second magnetic star showing the presence of a CIR, suggesting that WRs with 
CIRs are good candidates for future magnetic surveys.

Notably, pulsations cannot cause the short periodicity in WR\,55, as we see 
a change in field polarity. While low-amplitude magnetic variability has 
been found in pulsating magnetic stars, these changes in $\bz$ are 
$\sim10-15$\,G 
\citep[e.g.,][]{Shultz}.
Similar short periodicities have been reported for other stars
(e.g., 9.8\,h in WR\,123, 
\citealt{Chene};
15.5\,h in WR\,46, 
\citealt{Henault-Brunet}).
With our period and a radius estimate of 5.23\rsun{} from
\citet{Hamann2019}, 
we find $v_{\rm eq}=534$\,\kms, suggesting that WR\,55 was perhaps spun up 
by binary interaction
\citep{deMink}.
Magnetic spectropolarimetry of WR stars, especially nitrogen-sequence WN 
stars, is crucial, as they are likely progenitors of type Ib or IIb 
supernovae 
\citep[][]{Stevance}.

Finally, we compare our photometric results for WR\,55 with the large samples of
\citet{naze-2021} 
and 
\citet{lenoir-craig-2022}.
Table 2 of 
\citet{lenoir-craig-2022}
shows their best-fit values of the four semi-Lorentzian parameters; we focus 
on WR\,78 and WR\,87, which are like WR\,55: WN-type stars with both 30- and 
2-minute cadence data. These two stars have $A_0$ and $C_W$ decreasing between
sectors, but $\nu_{\rm char}$ and $\gamma$ increase, which agrees with our 
findings for WR\,55. This effect (especially for $A_0$ and $C_W$) may partially
arise from the differing cadences and noise properties of the data; this is 
explored in fig.~5 of 
\citet{lenoir-craig-2022}.
However, $\gamma$ (and $\nu_{\rm char}$) correspond to intrinsic properties
of the star, so their change across sectors is interesting and may indicate
the existence of slow, internal stellar processes. For instance, both
our values of $\gamma$ agree with the predictions of
\citet{2019ApJ...876....4E}
and may indicate core-generated internal gravity waves. More data will help
untangle the relative effects of sources of flux variability.

\section*{Acknowledgements}
We thank our referee, Dr.\ Pascal Petit, for insightful comments.
This work is based on observations made with ESO telescopes at the La Silla 
Paranal Observatory under programme IDs 0104.D-0246(A) and 109.230H.001.
This paper includes data collected by the TESS mission. Funding for the TESS 
mission is provided by the NASA Science Mission Directorate. Resources 
supporting this work were also provided by the NASA High-End Computing (HEC) 
Program through the NASA Advanced Supercomputing (NAS) Division at Ames 
Research Center to produce the SPOC data products 
\citep{jenkins-spoc}. 
The 2-min cadence data of WR\,55 were obtained via Guest Investigator
program G03095 (PI: T.\ Dorn-Wallenstein).

\section*{Data Availability}

All FORS\,2 data are available from the ESO Science Archive Facility at
\url{http://archive.eso.org/cms.html.} 

TESS data can be downloaded from the 
Barbara A.\ Mikulski Archive for Space 
Telescopes (\url{mast.stsci.edu}).

\bsp	
\label{lastpage}
\end{document}